\documentstyle[preprint,revtex,eqsecnum]{aps}

\begin{document}

\draft
\preprint{U. of MD PP \#94-062}
\preprint{DOE/ER/40762-022}
\preprint{WM-93-110}
\preprint{December 1993}
\begin{title} Systematics of Exclusive $B$\/--\/decays
\end{title}

\author{C. E. Carlson}
\begin{instit} Physics Department, College of William and Mary\\
Williamsburg, VA 23187, USA\
\end{instit}

\author{J. Milana}
\begin{instit} Department of Physics, University of Maryland\\
College Park, Maryland 20742, USA\\
\end{instit}

\begin{abstract}
  Within a perturbative QCD framework we present a systematic
analysis of the high recoil,  in particular two body mesonic, decay
channels of the $B$-meson.  For decays involving
$D$-mesons, all experimentally observed rates are in reasonable
accord with our calculations.  Noticeable corrections to heavy
quark symmetry are found, appearing most prominently when
comparing $D$ and
$D^*$ decay modes.   Calculated branching ratios involving the
CKM matrix element
$V_{ub}$ are all below experimental upper bounds, but not
markedly so.  Calculated decay rates into charmonium lie a bit on
the low side.  Where color-suppressed diagrams contribute,  they
are found to be also suppressed dynamically  and explicit
calculation shows that the color-suppressed diagrams are ignorable
in $B
\rightarrow D \pi$ decays.
\end{abstract}

\newpage
\narrowtext
\section{Introduction}

The two body exclusive decay modes of the $B$-meson are short
distance events and can thus be addressed by perturbative QCD
(pQCD).  In an earlier letter \cite{Bdec}, building from the work of
Szczepaniak {\it et al.} \cite{Adam}, we developed a formalism for
two body mesonic decays and applied it to three illustrative cases:
$B^0
\rightarrow D^- D_s^+$, $B^0 \rightarrow D^-
\pi^+$ and $B^0 \rightarrow \pi^- \pi^+$.  In this follow up work,
we present a systematic analysis of the two body mesonic decays of
the
$B$-meson,

\begin{equation} B \rightarrow m_1 + m_2,\label{generic}
\end{equation}  where $m_1$ and $m_2$ are S-wave mesons and
most often ground states for their respective quantum numbers.
There are many such decays.

It is convenient to separate the decays (\ref{generic}) into two
sets, according to the CKM matrix element entering the decay:
either $V_{cb}$ or $V_{ub}$. The
$V_{ub}$ set contains only decays that are as yet unseen
\cite{PDG}, and includes all the light-light meson decay modes as
well as some specific light-heavy decay channels, such as
$B^0 \rightarrow \pi^- D_s^{+}$.  The
$V_{cb}$ set can be further divided into those involving
$D$-mesons, and those involving charmonium.

The modes involving $D$-mesons,  to the extent they are factorable
at the hadronic level (see below if a definition is needed),  depend
upon form factors which can all be related to a single form factor
using a global symmetry among heavy flavors of quarks in QCD.
This symmetry is well associated with the work of Isgur and Wise
\cite{WG},  although we should also note the early work of
Nussinov and Wetzel
\cite{nussinov}. While heavy quark symmetry can be applied for
any momentum transfer, it is expected to be most accurate near the
zero recoil point. This is the point where the $B$ and $D$ have zero
relative velocity, or
$v\cdot v^\prime =1$, where $v$ and $v^\prime$ are the
four-velocities of the two heavy mesons. At the zero recoil point,
the single form factor
$\xi(v \cdot v^\prime)$ takes the known value
$\xi(1)\approx1$.  We however are here concerned with decays
nearly as far from the zero recoil point as is kinematically
possible.  It should not therefore be surprising that we find large
corrections to heavy quark symmetry,  and we may give three
reasons why large corrections are not surprising.

First, explicitly ignoring violations of heavy quark symmetry,  one
can show from pQCD that $\xi(v \cdot v^\prime)$ is a dipole
\cite{Bdec,KandK} at large $\omega = v \cdot v^\prime$.  The dipole
form is pleasing as it is also what one obtains from atomic physics
and thus confirms the common statement that the
$B$-meson is the hydrogen atom of QCD.  However, the dipole result
is inconsistent with the general pQCD counting rule result
\cite{scaling} that meson form factors fall like monopoles at large
$Q^2$. Indeed this is what one finds in complete pQCD calculations;
the additional momentum dependence comes from terms in the
numerators that could be neglected for infinite heavy quark
mass.    Large corrections to heavy quark symmetry should thus be
expected  in, for example, $e^+e^-$ annihilation into heavy
mesons,  where one has
$Q^2 \geq m^2_{H}$ and $\Lambda Q^2/m^2$ corrections to the
numerator of
$\xi(\omega)$ can dominate the physical form factor
\cite{eeprod}.  At the momentum transfers involved in
$B$-decay,  the heavy quark symmetry violating terms are not
dominant but are noticeable.

Second, heavy quark symmetry applied to the meson decay
constant leads to
$f_H \propto 1/\sqrt{m_H}$.  It is necessary
\cite{eeprod} to use this scaling of the meson decay constant in
order to obtain universality of the Isgur-Wise form factor at large
recoil for different heavy mesons.  Recent lattice calculations
\cite{lattice}, however, find that the meson decay constant is
nearly the same for the $D$ and $B$ mesons,  suggesting large
violations of heavy quark symmetry predictions for $f_H$.  If this
lattice result is correct, the explicit connection made by pQCD
predicts correspondingly large violations of heavy quark
symmetry in
$\xi(\omega)$ for large recoil.

Last, and most easily seen in the present work, $\xi(\omega)$ is a
sensitive function of a light quark mass and binding energy scale
$\Lambda^\prime$ (explicitly, the dependence is
$\Lambda^{\prime {-3}}$).  The sensitivity comes from propagator
denominators,  and large corrections are possible because the
denominators can also be sensitive to mass differences (e.g.,
between the $D$ and $D^*$) that are known to be sizable on the
light quark mass scale.  There are also smaller corrections from
numerator terms.  The fact that
$\xi(\omega)$ and hence the calculated branching ratios are
sensitive functions of the soft physics parameters or quantities
does not vitiate the applicability of pQCD.  Recall that hard inclusive
cross-sections have sensitive dependence upon soft physics via the
structure functions.  The crucial issue here, as there, is that the
process being considered be a short distance event so that a twist
expansion (an expansion in inverse powers of some large
momentum or mass squared) is justifiable.  The twist expansion is
indeed justifiable for the two body decays of a $B$-meson where the
energy of the initial state ($M^2 \approx 28$GeV$^2$) is being
carried away by two highly Lorentz contracted, effectively
massless, mesons.

For the two body mesonic decays involving either charmonium
production or the CKM matrix element $V_{ub}$, an important
result of our calculations is that the decay amplitude acquires an
imaginary part.  This imaginary part appears because some heavy
quark propagators in some Feynman graphs can go on-shell in the
integration over the mesonic distribution amplitudes.  Picking up
such poles is legitimate in pQCD
\cite{GlennySterman,Kahler} as they are not pinched singularites
and hence, by the Coleman--Norton theorem
\cite{CNorton}, are not associated with a long distance interaction.
The existence of these imaginary parts are crucial in obtaining
rates for the charmonium decays comparable to the observed
values.  In the case of the unobserved light-light meson decay
channels, they help place our estimates just below the
experimental upper bounds.

The organization of the rest of this paper is as follows. Section II
discusses the decays proportional to the CKM matrix element
$V_{cb}$ which involve $D$-mesons.  As all our theoretical
apparatus is presented in this section, it itself is divided into
several parts: (a)~discusses our general formalism and presents our
results for the heavy meson transition amplitudes when ignoring
all `color-suppressed' diagrams; (b)~discusses our expressions in
the heavy quark symmetry limit; (c)~presents an explicit
calculation of the color-suppressed diagrams in the case of $B
\rightarrow D \pi$ decays, demonstrating that they are in fact
ignorable and also introducing the key ingredient for later
sections that the hard-scattering amplitudes can acquire
imaginary parts; (d)~catalogs our results with all terms in place
and compares to data, and also shows the deviations from heavy
quark symmetry at high recoil; and lastly, (e)~details an potentially
useful interpolating model whose form factors fitted our pQCD
results at high recoil and the heavy quark symmetry results at zero
recoil. Section III presents our results for decays involving
charmonium while section IV does likewise for decays proportional
to $V_{ub}$. Section V contains our general conclusions and a
discussion of future prospects.

\vfill \eject

\section{The Heavy Meson Transitions}

\subsection{{\it General formalism and the Heavy Meson transition
matrix
 elements}}

Exclusive processes at large momentum transfer are addressed
\cite{BrodLep} within pQCD starting with a Fock component
expansion of the involved hadrons.  A twist expansion suggests that
only the contribution from the lowest order Fock component
should dominate the physical observable under consideration
\cite{hadhad}.  For mesons, this simply becomes the
quark--antiquark Fock component.  An exclusive process then
involves a perturbatively calculable hard amplitude convoluted
with a nonperturbative, soft physics wave function,
$\psi_m$, from each of the hadrons $m$ entering or leaving the
hard interaction.  For our calculations we employ the factorization
scheme advocated by Brodsky and Lepage
\cite{BrodLep}, and take the momenta of the quarks as some
fraction $x$ of the total momentum of the parent meson weighted
by a soft physics distribution amplitude $\phi(x)$ ($\phi(x)$ being
then simply the quark's wave function $\psi(x,k_\perp)$
integrated over transverse momentum, $k_\perp$)
\cite{comment1}.  In the case of a pseudoscalar meson, the
normalization of the two quark Fock component distribution
amplitude is determined by it's leptonic decay matrix element,
shown in Fig. \ref{decay}.  For the $B$ and $D$ mesons this
becomes,

\begin{equation}
\left \langle 0 \left| V^\mu - A^\mu \right | H^+ \right
\rangle = - 3i
\frac{ {\rm Tr}[ \gamma^5 (\not\!\kern-1pt p_{H}-m_{H})
\gamma^\mu ( 1 -
\gamma^5 ) ]}{\sqrt{2}\sqrt{3}} \int_{0}^{1} dx
\phi_{H}(x)\label{fdecay} \end{equation} where the factor
${\gamma^5 (\not\!\kern-1pt p_{H}-m_{H})}$ arises from the sum
over quark spinors constrained to be in a S--wave, pseudoscalar
state; the overall minus sign corresponds to the presence of a
fermionic loop; the 3 in the numerator arises from a sum over
quark colors; and the factors in the denominator are for
normalization purposes.  The relation

\begin{equation}
\left \langle 0 \left | A^\mu \right | H^+ \right \rangle = i
\sqrt{2} f_{H} p_{H}^\mu,
\end{equation} is obtained using the normalization condition

\begin{equation}
\int_{0}^{1} dx_1 \phi_{H}(x_1) =  \frac{1}{2 \sqrt{3}} f_{H}.
\end{equation}

In the case of the heavy mesons, it is expected that the momentum
fraction of the light quark, $x_1$, is small and that
$\phi(x_1)$ is a strongly peaked function about some average
value $\epsilon_{B,D}$. For simplicity, we have therefore used that
\begin{equation}
\phi_{B,D}(x_1) = \frac{1}{2 \sqrt{3}} f_{B,D} \delta (x_1 -
\epsilon_{B,D}).\label{heavydist}
\end{equation} The parameter $\epsilon$ we relate to the mass
difference
$\Lambda$ between the heavy meson and its constituent heavy
quark
\begin{equation} m_{B,D} = m_{b,c} + \Lambda_{B,D}\label{deflam}
\end{equation} whereby one obtains the relation that
\begin{equation}
\epsilon_{B,D}= \frac{\Lambda_{B,D}}{m_{B,D}}.\label{defeps}
\end{equation}

The lowest order Feynman diagrams that contribute to the two body
mesonic decay of the B meson are shown in Figs.\
\ref{mesdec},
\ref{cs1}, and \ref{cs2}.  Figures \ref{mesdec}(a) and
\ref{mesdec}(b), where the $W$ boson decays directly into one of
the final state mesons of mass $q^2$, are factorable at the hadronic
level,  and in some literature are called ``color-unsuppressed"
diagrams as they contain an explicit
$N_c=3$ enhancement over figures
\ref{cs1} and \ref{cs2} \cite{comment2}.  In subsection (IIc) we
show that in the explicit case of $B \rightarrow D \pi$ decays these
latter diagrams are in fact further suppressed due to dynamical
considerations.  We will hence ignore them now and focus only on
the contributions from Figures
\ref{mesdec}(a) and \ref{mesdec}(b).  The relevant amplitude
entering the decay of the B meson then involves the transition
matrix element
\begin{equation}
\left \langle D^- \Big | A^\mu \Big | B^0 \right \rangle =
\int_{0}^{1} dx dy \, \phi_B (x) \phi_D (y) (t^\mu_1 +
t^\mu_2)=T_1^{\mu}+T_2^{\mu},\label{transdef}
\end{equation} where $t^\mu_1$ and $t^\mu_2$ are the
corresponding hard amplitudes to Figures \ref{mesdec}(a) and
\ref{mesdec}(b) respectively.  Using Eqs.\
\ref{heavydist}--\ref{defeps}, one obtains that
\begin{eqnarray} T_1^{\mu} &=& \frac{16 \pi \alpha_s f_B f_D}{9
D_0 D_1}
\, \Big[ p_D^\mu m_B \Big(m_B - 2 \Lambda_B + 2 \Lambda_D
(\omega-1)\Big) + p_B^\mu m_D (m_B +
\Lambda_B - \Lambda_D) \Big],\nonumber\\ T_2^{\mu} &=&
\frac{16
\pi
\alpha_s f_B f_D}{9 D_0 D_2} \, \Big[ p_D^\mu m_B (m_D +
\Lambda_D -
\Lambda_B) + p_B^\mu m_D \Big(m_D - 2
\Lambda_D + \Lambda_B (\omega-1)\Big)\Big],\label{hardBD}
\end{eqnarray} where the denominators, corresponding to the
various propagators, are
\begin{eqnarray} D_1 =& (y p_D - p_B)^2 - m^2_b + i\eta =& 2 m_B
(\Lambda_B -
\Lambda_D \omega)\nonumber\\ D_2 =& (x p_B - p_D)^2 - m^2_c +
i\eta =& 2 m_D (\Lambda_D -
\Lambda_B \omega)\nonumber\\ D_0 =& (x p_B - y p_D)^2 + i\eta =&
-2\Lambda_B \Lambda_D
\omega.\label{denom} \end{eqnarray} As defined in the
introduction,
\begin{equation}
\omega = v \cdot v^\prime = \frac{m_B^2 + m_D^2 - q^2}{2 m_B
m_D},
\end{equation} and is in the case of the two body exclusive decays
as far from the zero--recoil point as is kinematically allowable.

In the above expressions, we have dropped all terms in the hard
amplitude of order $\epsilon_{B,D}^2$.   These are not only expected
to be numerically small but are related, through the mass of the
light quark, to transverse momentum effects.  Since we ignore the
latter in our factorization scheme, self--consistency in fact dictates
that these other terms also be ignored.

For the heavy meson transitions of the B into a D$^*$,
\begin{equation}
\left \langle D^{*-} \Big | V^\mu - A^\mu \Big | B^0 \right
\rangle =
\int_{0}^{1} dx dy \, \phi_B (x) \phi_{D^*} (y) (t^{*\mu}_1 +
t^{*\mu}_2) = T^{*\mu}_1+ T^{*\mu}_2,\label{transtardef}
\end{equation} where now $T^{*\mu}_1$ and $T^{*\mu}_2$ are
\begin{eqnarray} T_1^{*\mu} =& \frac{16 \pi \alpha_s f_B
f_{D^*}}{9 D^*_0 D^*_1}\, \Big[ &\epsilon^\mu m_B m_{D^*}
\Big(m_B(\omega^*+1) +   (
\Lambda_B + \Lambda_{D^*})(\omega^*-2)\Big) \nonumber\\
 &&- p_{D^*}^\mu \epsilon \cdot p_B (m_B +
\Lambda_B - \Lambda_{D^*})\nonumber\\
 &&+ i \epsilon_{\mu\nu \alpha \beta } \, \epsilon^\nu
p^\alpha_{D^*}  p^\beta_B \, (m_B + \Lambda_B - \Lambda_{D^*})
\Big]\nonumber\\ T_2^{*\mu} =& \frac{16 \pi \alpha_s f_B
f_{D^*}}{9 D^*_0 D^*_2}\, m_{D^*}\,
\Big[ &\epsilon^\mu m_B \Big(m_{D^*}(\omega^*+1) - \Lambda_B -
\Lambda_{D^*}\Big) \nonumber\\
 &&- {p_{D^*}^\mu} {\epsilon \cdot p_B} + 2 \epsilon_B p_B^\mu
\epsilon
\cdot p_B \nonumber\\
 &&+ i \epsilon_{\mu\nu \alpha \beta } \, \epsilon^\nu
p^\alpha_{D^*}  p^\beta_B \Big]\, ,\label{hardBDstar}
\end{eqnarray} in which $\epsilon$ is the polarization vector of
the
$D^*$.  The decay constant of a vector meson is defined in general
by

\begin{equation}
\left \langle 0 \left | V^\mu \right | V^+ \right \rangle =
\sqrt{2} f_V m_V
\epsilon^\mu \hskip8pt,  \label{vectordecayconstant}
\end{equation} where $\epsilon$ is the polarization vector of the
vector meson.  With these conventions,  $f_V$ and $f_P$ would be
the same if $V$ and $P$ were corresponding heavy meson states,
but for the $\rho$ and
$\pi$ the experimental values are $f_{\rho}$ = 151 MeV and
$f_{\pi}$ = 93 MeV.

The parameters $\omega^*,\Lambda_{D^*}$, and $\epsilon_{D^*}$,
and the denominators$D^*_{0,1,2}$ are defined as in the previous
$B$ to $D$ transition case except to replace the $D$ mass with the
$D^*$ mass.  As mentioned in the introduction and as we will see
further in section (IID), these mass differences should not be
neglected when obtaining physical decay rates. To illustrate their
importance, however, we first discuss our results when they are
ignored, obtaining the Heavy Quark Symmetry limit of our pQCD
results.

\subsection{{\it Wisgur at large recoil}}

\nopagebreak The Heavy Quark Symmetry limit of our results can
be obtained in a simple two step procedure.  First, take
$\Lambda$ (which is setting the scale of the soft physics) to be
universal for all heavy mesons (and hence, implicitly setting
$m_D=m_{D^*}$).  Second, consistently ignore all
$\Lambda/M$ corrections compared to terms of
$O(\Lambda/M)^0$. From Eq.\ \ref{hardBD} one then obtains that
(in the notation of \cite{WG})
\begin{eqnarray}
\left \langle D^- \left | A^\mu \right | B^0 \right \rangle &=& f_+
(p_B + p_D)^\mu + f_- (p_B - p_D)^\mu\nonumber\\ f_+ &=&
-\xi_+(\omega) (m_B + m_D) \big/ \sqrt{4m_Bm_D} \nonumber\\
f_- &=& \xi_-(\omega) (m_B - m_D)
\big/ \sqrt{4m_Bm_D} \label{wisgur1}
\end{eqnarray} where for exact heavy quark symmetry,
$\xi_{\pm}(\omega) =
\xi(\omega)$, which is the Isgur--Wise form factor and which is
given in pQCD to be at large $\omega$

\begin{equation} {\xi(\omega) \over \sqrt{4m_Bm_D}} = -
\frac{4\pi
\alpha_s f_B f_D}{9 \Lambda^3 \omega (\omega - 1)}.\label{wisff}
\end{equation} As in the case of the hydrogen atom in QED,
$\xi(\omega)$ is a dipole. Note though that this is only true for
what might be called `moderately'--large $\omega$.  At
`very'--large $\omega$, as occurs in
$e^+ e^-$ annihilation \cite{eeprod} and in which the mass of the
heavy meson is no longer the largest scale in the problem, some
$\Lambda/M$ ``corrections" in the numerator may be multiplied by
factors of $\omega$ and then cannot be ignored.  These
``corrections" ultimately dominate the form factor, reproducing
thereby the canonical pQCD scaling result
\cite{scaling} that all meson form factors behave like $1/Q^2$.

For the case of the $B$ to $D^*$ transitions, one obtains from Eq.\
\ref{hardBDstar} that
\begin{eqnarray}
\left \langle D^{*-} \Big | V^\mu - A^\mu \Big | B^0 \right
\rangle &=& -ig\epsilon_{\mu \alpha \beta \gamma}
\epsilon_\perp^\alpha (p_B - p_{D^*})^\beta (p_B +
p_{D^*})^\gamma \nonumber\\ &&-f\epsilon^\mu - a_+\epsilon
\cdot p_B (p_B + p_{D^*})^\mu - a_-\epsilon \cdot p_B (p_B -
p_{D^*})^\mu, \nonumber\\ f &=& \xi_f^*(\omega) 2 m_B m_{D^*}
(1 + \omega)
\big/ \sqrt{4m_Bm_D^*},
\nonumber\\
 g &=& \xi_g^*(\omega) \big/ \sqrt{4m_Bm_D^*},  \nonumber\\
 a_\pm &=& {\mp} \xi_\pm^*(\omega) \big/ \sqrt{4m_Bm_D^*},
\label{wisgur2} \end{eqnarray} in which the heavy quark result
is now

\begin{equation}
\xi_f^*(\omega) = \xi_g^*(\omega) = \xi_\pm^*(\omega) =
\xi^*(\omega) =
\xi(\omega) \frac{f_{D^*}}{f_D}.\label{wisffp} \end{equation}

The general result of Heavy Quark symmetry that the form factors
Eq.\
\ref{wisff} and \ref{wisffp} are universal functions then follows
provided that
\begin{equation} f_M \propto \frac{1}{\sqrt{M}}.\label{hdecay}
\end{equation} As Eq.\ \ref{hdecay} is itself a standard Heavy
Quark result, Heavy Quark Symmetry and pQCD are,  in principle, in
perfect accord.  On the other hand, the results for the meson decay
constant and the Isgur--Wise form factor at large recoil are
intimately related.  This should not be unexpected as they are both
related to the short distance (i.e. the two quark Fock component)
structure of the heavy meson.  A nontrivial result of this
connection is that any violations in one must appear in the other.
This fact has particular relevance considering the recent lattice
\cite{lattice} results which indicate that  $f_B$  and
$f_D$ are in fact comparable,  contrary to \ref{hdecay}.

In closing, observe finally that $\xi(\omega)$, Eq.\
\ref{wisff}, depends upon $\Lambda^{-3}$.  An actual decay rate,
being related to
$\xi^2$, is therefore proportional to $\Lambda^{-6}$.  Two
body--exclusive meson decays are thus highly sensitive to
violations to heavy quark symmetry.  In particular and as we will
soon see, the $D-D^*$ mass difference cannot be ignored.  Before
though demonstrating this point, we wish to first justify our
approximations above of keeping only the color-unsuppressed
diagrams by an explicit evaluation of the color-suppressed
diagrams, figures \ref{cs1} and
\ref{cs2}, for the $B \rightarrow D \pi$ decay channels.  These
latter are of particular interest as the latest experimental quotes
\cite{Bdata} suggest large differences between the $B^-
\rightarrow D^0 \pi^-$ and $B^0 \rightarrow D^+ \pi^-$
branchings.

\subsection{{\it $B \rightarrow D \pi$: Color suppression
dynamically enforced}}

What we called factorable diagrams in \cite{Bdec} are those that
can be written at the hadronic level as a matrix element of one
weak current multiplied by the matrix element of another current.
For the present example,  this means a matrix element of a weak
current for the transition
$B \rightarrow D$ (or
$B
\rightarrow D^*$) multiplied by the matrix element of a weak axial
current between the vacuum and the pion.  At the quark level,  this
means that the two quarks that come from the decay of the virtual
$W$ both go into the same hadron,  the pion in the present
example.

There are other possibilities.  One of the quarks from the $W$ decay
may pair with the spectator quark to form one hadron,  and the
other quark from the
$W$ decay may pair with the charm quark coming directly from
the $b$-quark decay to form the other hadron.  See Figs.
\ref{cs1} and \ref{cs2}.  As has been pointed out,  these ``internal
$W$'' or ``$W$-exchange'' diagrams are suppressed by color factors
of $O(N_c)$.  (The factorable diagrams have two independent color
loops and the non-factorable diagrams have only one.)  In fact,  a
similar argument applies to the spin sums,  so that one may argue
that the suppression is actually of $O(2 N_c)$.   However,  this is
only a rough argument since the dynamics of the two classes of
diagrams are different.  When we do the full calculation,  as we are
about to do,  we discover that the suppression of the color
suppressed diagrams is significantly greater than suggested by the
simple estimates just cited.

The leading order, color--suppressed diagrams are internal $W$ for
$B^+
\rightarrow D^0 \pi^+$ and are shown in Fig. \ref{cs1}; those for
$B^0\rightarrow D^- \pi^+$ are $W$-exchange and are shown in
Fig.
\ref{cs2}.  In addition to showing the full extent of the suppression
of the color suppressed diagrams,  the evaluation of these diagrams
introduces an important feature for our calculations in later
sections, and for clarity we will be rather detailed in our
presentation.

We use the $B^-$ decay for illustration.  The amplitude ${ M}$ is the
sum of contributions from the four diagrams of Fig.
\ref{cs1}, which yield the following expressions

\begin{eqnarray} {M}_a &=& {A\over\epsilon_B} \int_0^1 {dx \,
\overline{x}\,{\tilde{\phi}}_\pi(x)}
\frac{m_B^2 (1 - 2\epsilon_B)  + x (m_B^2 - m_D^2)}{x(m_B^2 -
m_D^2) - 2
\epsilon_B m_B^2 - i\eta} \nonumber\\ {M}_b &=&
-{A\over\epsilon_B}
\int_0^1 {dx \, \overline{x}\,{\tilde{\phi}}_\pi(x)}
\frac{m_B^2 (1 - \epsilon_D - \epsilon_B) -\epsilon_B m_D^2}{x(1
-\epsilon_D -\epsilon_B)(m^2_B - m^2_D) - {\epsilon_B (1 -
\epsilon_D) (m_B^2 + m_D^2)} + i\eta}\nonumber\\  {M}_c &=&
{A\over\epsilon_B}\, {m_D^2\over{m_B^2-m_D^2}}
\int_0^1 {dx \, \overline{x} \, {\tilde{\phi}}_\pi(x)}\nonumber\\
{M}_d &=& -{A\over\epsilon_B} \int_0^1 {dx \, \overline{x} \,
{\tilde{\phi}}_\pi(x)}
\frac{m^2_B(2\epsilon_B-x-\epsilon_D) +
m^2_D(x-\epsilon_D)}{x(\epsilon_D - \epsilon_B)(m^2_B - m^2_D) -
{\epsilon_D\epsilon_B(m^2_B + m^2_D)} +
i\eta},\label{diagramscs1}
\end{eqnarray} in which $\overline{x}=1-x$ and

\begin{equation} A={16\pi \over {9}} \alpha_s G_F V_{cb} V_{ud}
f_B f_D f_\pi\hskip6pt,
\end{equation} with $G_F$ being Fermi's constant, and $V_{cb}$
and
$V_{ud}$ are the appropriate CKM matrix elements.  The amplitude
is normalized so that the decay rate is given by

\begin{equation}
\Gamma ={\triangle(m_B^2,m_D^2,m_\pi^2) \over 16 \pi m_B^3}
\left| {M}\right|^2 \hskip6pt,
\end{equation} where $\triangle$ is the K\"all\'en triangle
function

\begin{equation}
\triangle (x,y,z) = x^2 + y^2 +z^2 - 2xy -2yz - 2zx.\label{triangle}
\end{equation}

Observe that except in ${M}_c$, all of the integrands in Eq.\
\ref{diagramscs1} contain a pole at some of value of $x=x_i$ within
the integration limits of the pion's distribution amplitude, $0 < x_i <
1$. (The apparent pole at $x=0$ is a typical endpoint singularity
regulated by
$\phi_\pi(x)$ itself.)  These poles correspond to a kinematically
accessible situation of a heavy--quark propagator (either the $c$
or $b$ quark, as the case may be) going on--shell.  However,  these
are not pinched singularities and thus one is able to deform the
integration path away from the pole.  Consequently, from the
Coleman--Norton theorem
\cite{CNorton}, these do not correspond to long distance
propagations of a heavy quark and the diagrams are calculable
 in pQCD.

Using the usual replacement
\begin{equation}
\frac{1}{x - x_o \pm i\eta} = \frac{1}{x - x_o} \mp i\delta(x -
x_o)\hskip6pt,
\end{equation} we may evaluate each of the integrals and find the
color suppressed contributions to the decays.  Note that the
amplitude acquires an imaginary part.

We calculate results for both the asymptotic \cite{asympt} and the
Chernyak-Zhitnitsky \cite{CZpi} pion distribution amplitudes.  The
asymptotic distribution amplitude is

\begin{equation}
\phi_\pi(x) =  \sqrt{3} f_\pi x(1-x) {\tilde{\phi}}_\pi =
\sqrt{3} f_\pi x(1-x)
\hskip6pt,
\end{equation} while the Chernyak-Zhitnitsky distribution
amplitude is

\begin{equation}
\phi_\pi(x) =  \sqrt{3} f_\pi x(1-x) {\tilde{\phi}}_\pi= 5\sqrt{3}
f_\pi x(1-x) (2x-1)^2 \hskip6pt.
\end{equation} The Chernyak-Zhitnitsky distribution amplitude is
motivated by QCD sum rule calculations of a few moments of the
pion distribution amplitude.  When used in other pQCD calculations,
such as for the pion electromagnetic form factor at high
momentum transfer,  tends to give results in accord with the quoted
data \cite{Bebek} (see however
\cite{CM90}).  There are,  on the other hand,  those who believe
\cite{MR} that the asymptotic form is more correct even at current
experimental energies (and that ``non-perturbative'' processes are
responsible for the size of the pion form factor,  etc.,  at current
energies).

Our results for the W-exchange contributions (Fig. \ref{cs2}) to
$B^0
\rightarrow D^- \pi^+$ are

\begin{equation} { M} = \cases { (-22 +11 i)A, &asymptotic; \cr
(-20+27i)A, &CZ\hskip6pt.}
\end{equation} However,  the standard of comparison is ${ M} = 785
A$ from the factorable graphs, so that the $W$-exchange
contribution is seen to be small.  Incidentally,  the
$W$-exchange gives the same contribution to
$B^0
\rightarrow D^0 \pi^0$ as it does to $B^0 \rightarrow D^-
\pi^+$,  but since there is no dominant factorable graph in the
former case,  its branching ratio is tiny.

For the charged $B$ decay $B^+ \rightarrow D^0 \pi^+$,  the
internal
$W$ graphs give

\begin{equation}  { M} = \cases { (61 + 17i)A, &asymptotic;\cr
(-25+57i)A,&CZ\hskip6pt,}
\end{equation} which are notably larger than in the color
suppressed contribution to the neutral $B$ decay.  The main effect
comes from interference with the factorable graph,  and the
asymptotic case leads to a roughly 15\% increase of the calculated
value of the branching ratio,  whereas the Chernyak-Zhitnitsky
case gives a few percent decrease.  These effects are again not big,
particularly if one favors the Chernyak-Zhitnitsky case.  For
comparison,  Bauer, Stech, and Wirbel
\cite{BSW}, for the same $D \pi$ final states we are discussing,
have the neutral $B$ decay about 50\% larger than the charged $B$
decay due to contributions of color suppressed diagrams.  Their
parameter setting the size of the color suppressed diagrams is
determined from the $B
\rightarrow J/\psi + K$ branching ratio;  we shall see later how we
deal with this decay.

\subsection{{\it The full factorable case}}

	Heavy quark symmetry is violated at the outset by the inequality
of the $D$ and $D^*$ (or $B$ and $B^*$) masses and additionally by
the inequality
$m_{B^*}-m_B\not=m_{D^*}-m_D$.  We deal with this,  and see the
consequences of it,   by thinking of the heavy meson mass as
having a piece that is due to the heavy quark mass,  a piece due to
binding which is the same for all heavy mesons,   and a piece due to
the spin-spin interaction.  Thus, for example,  we have
$m_{D^{(*)}} = m_c +
\Lambda_{D^{(*)}}$ and

\begin{eqnarray}
\Lambda_D = \Lambda' - 0.75 (m_{D^*}-m_D)\nonumber\\
\Lambda_{D^*} = \Lambda' + 0.25 (m_{D^*}-m_D) .
\end{eqnarray}

Corresponding equations hold for $m_{B^{(*)}}$,  with the same
$\Lambda'$.

	The calculations follow what was reported in subsection A,  and we
quote results in terms of the form factors defined in subsection B.
The form factors are of use for the semileptonic decays of the
$B$ as well as for the factorable two--body decays.  For the $B
\rightarrow D$ current we get

\begin{eqnarray} {f_\pm}(q^2) = {{8\pi}\over9}&&{{\alpha_s f_B
f_D} \over D_0}  \bigg\{ {{m_B}\over
   D_1}  \left[\pm m_B+m_D+{{m_D} \over m_B}(\Lambda_B -
\Lambda_D) \pm
   2(\Lambda_D (\omega-1)-\Lambda_B) \right] \nonumber\\
   + {{m_D}\over D_2} &&\left[\pm m_B+m_D \pm {{m_B} \over
   m_D}(\Lambda_D-\Lambda_B) + 2(\Lambda_B
(\omega-1)-\Lambda_D)
   \right]  \bigg\}
\end{eqnarray} where the denominators $D_0$,  $D_1$,  and $D_2$
are the same as before (Eq. \ref{denom}).

The results for the four form factors of the $B \rightarrow D^*$
current are

\begin{eqnarray} f(q^2) = {{8\pi}\over9} {{\alpha_s f_B f_{D^*}}
\over D_0^*} && 2m_Bm_{D^*} (\omega^*+1)
\Bigg\{{{m_B}\over
   D_1^*}  \left[1+{{(\Lambda_B + \Lambda_{D^*})(\omega^*-2)}
\over
  {m_B (\omega^*+1)}}\right]                   \nonumber\\
   && + {{m_{D^*}}\over D_2^*} \left[1-{{\Lambda_B +
\Lambda_{D^*}}
\over {m_{D^*}(\omega^*+1)}}  \right] \Bigg\}\hskip12pt,
\end{eqnarray}

\begin{equation} g = {{8\pi}\over9}{{\alpha_s f_B f_{D^*}}
\over D_0^*}
\Bigg\{{{m_B}\over
   D_1^*}  \left[1 + {{(\Lambda_B - \Lambda_{D^*})} \over
  {m_B }}\right]  + {{m_{D^*}}\over D_2^*} \Bigg\}\hskip12pt,
\end{equation}

\begin{equation} a_\pm = \mp {{8\pi}\over9}{{\alpha_s f_B
f_{D^*}} \over D_0^*}
\Bigg\{{{m_B}\over
   D_1^*}  \left[1 + {{(\Lambda_B - \Lambda_{D^*})} \over
  {m_B }}\right]  + {{m_{D^*}}\over D_2^*} \left[1 \mp
{{2\Lambda_B} \over
  {m_B}}  \right] \Bigg\}\hskip12pt,
\end{equation} where $D_0^*$,  etc., are got by appropriate mass
changes from the $B \rightarrow D$ case.

	With these form factors in hand,  the contribution of the
factorable graphs to the two-body decay rates yields,

\begin{equation}
\Gamma(B \rightarrow D P) = {{\Delta^{1/2}G_F^2 V_{cb}^2 V_{ij}^2
f_P^2}
   \over {16\pi m_B^3}} \left| (m_B^2 - m_D^2) f_+ + m_P^2 f_-
\right|^2
\hskip8pt,
\end{equation}

\begin{equation}
\Gamma(B \rightarrow D V) = {{\Delta^{3/2}G_F^2 V_{cb}^2 V_{ij}^2
f_V^2}
   \over {16\pi m_B^3}} \left| f_+ \right|^2 \hskip8pt,
\end{equation}

\begin{equation}
\Gamma(B \rightarrow D^* P) = {{\Delta^{3/2}G_F^2 V_{cb}^2
V_{ij}^2 f_P^2}
   \over {16\pi m_B^3}} {1\over{4 m_{D^*}^2}} \left| f + (m_B^2 -
   m_{D^*}^2) a_+ + m_P^2 a_- \right|^2  \hskip8pt,
\end{equation}

\begin{eqnarray}
\Gamma(B \rightarrow D^* V) &&={{\Delta^{1/2}G_F^2 V_{cb}^2
V_{ij}^2 f_V^2}
   \over {16\pi m_B^3}} {1\over{4 m_{D^*}^2}} \times \nonumber\\
  &&\left\{ \left|
  (m_B^2-m_{D^*}^2- m_V^2) f + \triangle a_+  \right|^2 + 8
m_{D^*}^2
  m_V^2  \left[\vert f \vert^2 + \vert g \vert^2 \right]
\right\}
\hskip2pt.
\end{eqnarray}

In the above,  $P$ and $V$ are pseudoscalar and vector mesons,
made from quarks $i$ and $j$ that correspond to the CKM factor
$V_{ij}$; each form factor has argument $m_P^2$ or $m_V^2$,
respectively; and
$\triangle$ is the K\"all\'{e}n triangle function defined earlier
with its arguments in each case being the masses squared of the
$B$ and of the two final state mesons.

 	We shall defer discussion of deviations from heavy quark
symmetry to the next section.  Eager readers may peruse the low
$q^2$ end of Fig. \ref{FF} or the high $\omega$ end of Fig. \ref
{FFom}.

  The results for the decay rates using the factorable diagrams are
shown in table \ref{table1},  in the column labeled ``basic model.''
Some comments should be made about the decay constants and
other parameters we used.

\begin{itemize}
\item We take our values of $f_B$ and $f_D$ from the central values
obtained from lattice gauge calculations,  in particular from Ref.
\cite{lattice}.  They obtain
$f_B = (1.42 \pm 0.52) f_\pi$  and $f_D = (1.58 \pm 0.49) f_\pi$ (we
summed their several error sources),  with a ratio determined with
smaller uncertainty,  ${f_B / f_D} = (0.90 \pm 0.05)$.  QCD sum rules
give results in the same vicinity;  for example Narison
\cite{Narison} reports
$f_B = (1.59 \pm 0.31) f_{\pi}$.

\item Errors in the choice of decay constants $f_B$ and $f_D$ are
compensated by small changes in our parameter $\Lambda'$.
Parameter
$\Lambda'$ itself must satisfy a criterion for reasonableness,
which we take to mean that it should be a few times
$\Lambda_{QCD}$,  and beyond that must give a good result for the
branching ratio of $B \rightarrow D
\pi$.  We use $\Lambda' = 545$ MeV\null.

\item We use $V_{cb} = 0.045$ and $V_{ud}=V_{cs}=0.975$,  and take
the $B$ lifetime to be 1.29 ps \cite{PDG}.

\item The value of $\alpha_s$ is taken as a geometric mean of the
values relevant for the off-shelledness of the quark and gluon
propagators.  We work simply to lowest order,  so have

\begin{equation}
\alpha_s = {4\pi \over \beta_0 \ln({Q^2 / {\Lambda_{QCD}^2}})}.
\end{equation}

Here $\beta_0 = 11 - (2/3)n_f$ and $n_f = 3$ is suitable for our
propagators.  We used $\Lambda_{QCD}=100$ MeV\null.  With the
value of
$\Lambda'$ quoted above and using the
$D \pi$ decay mode,  we get a value
$\alpha_s = 0.30$,  which we shall use for all our $B
\rightarrow D^{(*)}$ considerations.

\item We can determine the decay constant
$f_{D_s}$ from $B$ decay data,  since

\begin{equation}  {B(B\rightarrow D D_s) \over B(B \rightarrow D
\pi)} = 2.40 \left| f_{D_s} \over f_\pi \right|^2 \hskip8pt.
\end{equation}   This leads---for the basic model---to $f_{D_s}$
slightly greater than
$ f_\pi$,  with an uncertainty from the data of
$\pm 25\%$.  We have let $f_{D_s}= f_\pi$.  The lattice results
\cite{lattice} are larger,
$f_{D_s} = (1.74\pm 0.48) f_\pi$,  with $f_D / f_{D_s} = (0.90 \pm
0.05)$ (again).  Our ``model II'' below is interesting in this regard.

\item To obtain correct values for decays into
$D$ and $D^*$,  we cannot maintain the heavy quark symmetry
result of equality between these two meson's decay constants.
Rather,  we need something like $f_{D^*} = 1.39 f_D$.  Variations in
the values we use for
$f_B$ and $f_D$ do not greatly affect the calculated ratio of decay
rates for $B\rightarrow D \pi$ and $B\rightarrow D^*
\pi$.  Hence the ratio of the decay constants $f_{D^*}$ and
$f_D$ cannot be much different from the value just quoted (and
this statement stands even when we vary the model a bit in the
next section).  In particular,  we cannot maintain the heavy quark
symmetry result for these decays.

\item While the decay constant of the
$\rho$ is fixed from electromagnetic decays,  we have no
independent information that fixes the decay constant of the
$D_s^*$,  and we have {\it{pro tem}} set it equal to the decay
constant of the $D_s$.

\end{itemize}

Note that the decay $B \rightarrow D^* D_s^*$ is fairly copious,
although as yet unobserved.

\subsection{{\it An Interpolating Model}}

  Our calculations of the form factors are done using perturbative
QCD and are valid,   we believe,  for high momentum transfer in the
$B \rightarrow D$ and $B \rightarrow D^*$ currents.  Our
calculations are not valid near the no-recoil point or
pseudothreshold,  where the momentum squared of some of the
internal lines gets small and the process is not short distance.  In
fact,  some of the internal quark propagators go on shell
somewhere between the
$q^2 = 0$ point and the no-recoil point,  which leads to an infinity
in the form factor.  One notes by comparing to the form factors in
the $B
\rightarrow
\pi$ case below that this is an artifact of the peaking
approximation for the distribution amplitudes of the $D$ or
$D^*$,  but it not be fretted over as long as we are far enough from
the poles.  However,  one may wish for functional forms of the
form factors that interpolate between our calculated pQCD results at
the high momentum transfer end and the heavy quark results at
low momentum transfer.  We present such interpolating forms
here,   and refer to the resulting model as ``model II.''

  Good simple forms to use are dipoles,

\begin{eqnarray}  {\xi_{\pm}} = {\xi_{NR}  \left( 1 + {Q_-\over
{M_\pm}^2}\right)^{-2}}
\hskip6pt,
\nonumber\\ {\xi^*_i} = {\xi_{NR}  \left( 1 + {Q_-^*\over
{M_i^*}^2}
\right)^{-2}}
\hskip6pt,
\end{eqnarray} where $\xi_{NR} = 1.12$ is the value of the
Isgur-Wise function at the no-recoil point,  including some QCD
corrections,  and

\begin{equation}  Q_\pm^{(*)} = (m_B \pm m_D^{(*)})^2 - q^2 =
2m_B m_D^{(*)}(\omega^{(*)} \pm 1)
\end{equation} with $Q_-^{(*)}$ being zero at the no-recoil point.

We choose the masses $M_i^{(*)}$ to reproduce our results at
$q^2 = 0$.  The decay constants for $D$ and $D^*$ enter in our form
factor calculation and we again use $f_{D^*} = 1.47 f_D$. The masses
we get are listed for reference in table
\ref{table2}.

The ``interpolating model'' results for the six form factors are
shown graphically in Fig. \ref{FF}.  We plot the reduced form
factors,  which were defined in Section B and the equations for the
pQCD calculations may be read off from Section D.  Each of the
reduced form factors would be the Isgur-Wise function for exact
heavy quark symmetry.  As noted in the introduction,  we can
explicitly explore the violation of heavy quark symmetry in the
region where the pQCD results are valid.  The dispersion at $q^2 = 0$
shows that such violation is about 20\%.  Perhaps of some interest
are the same results plotted against
$\omega$ or $\omega^*$,  both of these being
$v\cdot v^\prime$.  This is shown in Fig. \ref{FFom}.  Four of the
six reduced form factors are in fact reasonably close,  but the other
two are significantly higher.  Incidentally,  the effect of just the
numerator corrections within the $B
\rightarrow D^*$ family may be seen by comparing to the reduced
form factor for $g$,  as the numerator corrections have little net
effect upon this form factor.

The results of ``model II'' for the branching ratios are listed in table
\ref{table1}.  There is a difference from the basic model in the
value determined for the decay constant
$f_{D_s}$,  which comes because the different behavior of the form
factors as a function of $q^2$ now gives

\begin{equation} {B(B\rightarrow D D_s) \over B(B \rightarrow D
\pi)} = 1.05 \left| f_{D_s} \over f_\pi \right|^2 \hskip8pt.
\end{equation}  which leads to $f_{D_s} \approx 1.7 f_\pi$.  We still
keep
$f_{D_s^*} = f_{D_s}$,  and we still find a copious $B
\rightarrow D^* D_s^*$ decay.

Although we have used some of the decay rates to determine
parameters (such as decay constants which could in principle be
determined elsewhere),  four of the eight branching ratios listed in
Table
\ref{table1} are still predictions in the sense that the parameters
required for their calculation are determined by other decays.
These four are the decays of the $B$ into
$D\rho$,  $D^* D_s$,  $D^* \rho$,  and the ratio of $D^* D_s^*$ to $D
D_s^*$.

\section{Decays involving Charmonium}

Decays of $B$-mesons into two body final states involving
charmonium proceed only via the ``internal-W'' or ``color
suppressed'' processes illustrated in Fig.
\ref{BJdecay}.  The final states we consider are those that are
leading in CKM factors and lowest S-state in spatial wave function.
This means we consider $B \rightarrow J/\psi+K$, $B
\rightarrow J/\psi+K^*$,  $B \rightarrow \eta_c+K$,  and $B
\rightarrow \eta_c+K^*$.  The amplitudes for
$B \rightarrow \eta_c + K$ are

\begin{eqnarray}   M_a &=&-{A \over \epsilon_B}\int_0^1dz \, \bar
z
\tilde \phi_K(z)
        {m_B^2(1-2\epsilon_B)+z(m_B^2-m_{\eta_c}^2) \over 2m_B^2
				\epsilon_B-z(m_B^2-m_{\eta_c}^2) + i\eta}  \nonumber \\
M_{bc} &=& {A \over
\epsilon_B}\int_0^1dz \, \bar z \tilde \phi_K(z)
      {2(m_B^2-m_{\eta_c}^2)(z-\epsilon_B) \over
z(1-2\epsilon_B)(m_B^2-m_{\eta_c}^2)-\epsilon_B(m_B^2+m_{\eta_c}^2)
+\epsilon_{\eta_c}m_{\eta_c}^2 + i\eta}
     \nonumber \\   M_{d} &=& {A \over \epsilon_B} \,  {m_{\eta_c}^2
\over m_B^2-m_{\eta_c}^2}  \int_0^1dz \,
     \bar z \tilde \phi_K(z) ,
\end{eqnarray}   where diagrams b and c are added together and
$\bar z = 1-z$.  We have allowed for the possibility $m_{\eta_c} =
2m_c+\Lambda_{\eta_c}$ and followed our usual practice in letting
$\epsilon_{\eta_c} = \Lambda_{\eta_c}/m_{\eta_c}$.  It is not clear
that we should use the same basic $\Lambda$ for the heavy-light
and heavy-heavy quark systems,  so that while one might consider
$\Lambda_{\eta_c} = \Lambda^\prime -
0.75(m_{J/\psi}-m_{\eta_c})$,  we have actually just set
$\Lambda_{\eta_c}$ to zero,  and correspondingly for the
$J/\psi$ later.  The $\Lambda$ parameter for charmonium appears
in only one term above and it is not the dominant term,  so that the
sensitivity to it is not great,  in contrast to the corresponding
parameters for the $B$ and $D$.  (An exception will be the $B
\rightarrow J/\psi+K^*_T$ given below where switching from
$\Lambda_{J/\psi} = 0$ to the other possibility given above will
increase that rate by about a factor three.)

The results for $B \rightarrow \eta_c + K^*$ are the same---only
the longitudinal $K^*$ is possible---except for using the $K^*$
distribution amplitude and making the usual change in $A$.

The formulas for $B \rightarrow J/\psi+K$ and $B \rightarrow
J/\psi+K^*_L$ ($L$ stands for longitudinal) are related to their
$\eta_c$ counterparts by simple mass and decay constant changes
and three more small changes,  which are the sign of the
$\epsilon_B m_{J/\psi}^2$ term in $M_{bc}$, the overall sign of
$M_d$,  and an additional term $+\epsilon_{J/\psi}m_{J/\psi}^2$ in
the numerator for $M_{bc}$. Incidentally,  much of the similarity
follows technically from
$m_{J/\psi} p_B \cdot \xi_L = p_B \cdot p_K$,  where $\xi_L$ is the
longitudinal polarization vector of the
$J/\psi$.  Also,  for the asymptotic wave function,
$\int dz \, \bar z \tilde \phi_K(z) = 1/2$.

For the $B \rightarrow J/\psi+K^*_T$ alone one has

\begin{eqnarray}
 M_a&=&-{A \over \epsilon_B}\int_0^1dz \, \bar z \tilde
\phi_{K_T^*}(z)
        {2 m_B m_{J/\psi}(1+\epsilon_B) \over 2m_B^2
				\epsilon_B-z(m_B^2-m_{J/\psi}^2) + i\eta}  \nonumber \\
M_{bc} &=& {A \over \epsilon_B}\int_0^1dz \, \bar z \tilde
\phi_{K_T^*}(z)
      {4m_B  m_{J/\psi}(2\epsilon_B - \epsilon_{J/\psi}) \over
z(1-2\epsilon_B)(m_B^2-m_{J/\psi}^2)-\epsilon_B(m_B^2+m_{J/\psi}^2)
     +\epsilon_{J/\psi}m^2_{J/\psi} + i\eta}  \nonumber \\  M_{d}
&=& 0 .
\end{eqnarray}  The constant $A$ for the $B \rightarrow
J/\psi+K$decay is

\begin{equation}   A={16\pi \over {9}} \alpha_s G_F V_{cb} V_{cs}
f_B f_{J/\psi} f_K\hskip6pt,
\end{equation}  and {\it{mutatis mutandi}} for the other decays.

In choosing the value of $\alpha_s$ to use in this case,  we should
anticipate what we will learn below,  which is that the imaginary
parts of the decay amplitudes are large enough to dominate the
decay rate.  Again,  having the quark on-shell does not vitiate the
perturbative expansion in this case,  but it does leave the gluon
propagator off-shellness as the only scale to use in estimating a
suitable argument for $\alpha_s$.  When the quark is on shell---we
take amplitude $M_a$ for definiteness;  the other possibility is not
so different---the gluon momentum squared is $2\Lambda_B^2$,
and continuing with $\Lambda_{QCD} = 100$ MeV,   we get
$\alpha_s = 0.35$.  We shall use this $\alpha_s$ throughout our $B$
decay into charmonium considerations.

We have for simplicity in this set of calculations set the mass of the
$K$ and $K^*$ to zero.  (Setting the mass of the $\rho$ to zero in
the $B
\rightarrow D \rho$ has only a small effect on the answer.)  For
our reaction mechanism,  no left handed $J/\psi$ or
$K^*$ will emerge in $B\rightarrow J/\psi+K^*$;  for the $\bar B$,
change hands.

We may use the asymptotic distribution amplitude for the $K$ or
$K^*$,  but in addition we quote results using kaon distribution
amplitudes of Chernyak, Zhitnitsky, and Zhitnitsky~\cite{CZZ},

\begin{eqnarray}
\phi_K(x) = \sqrt{3} f_K x_1 x_2 \left(3.0(x_1-x_2)^2 + 0.40
\right)
\hskip6pt,\nonumber\\
\phi_{K^*}(x) = \sqrt{3} f_{K^*} x_1 x_2 \left(0.5(x_1-x_2)^2 + 0.90
\right)
\hskip6pt.
\end{eqnarray}

We calculate supposing that the
$J/\psi$ is a non-relativistic bound state so that $p_c$,  the
momentum of the charmed quark entering the $J/\psi$,  is one-half
$p_{J/\psi}$.  The decay constant of the $J/\psi$ is got from its
electronic decay,

\begin{equation}
\Gamma(J/\psi \rightarrow e^+ e^-) = {32\pi \alpha^2 \over
27m_{J/\psi}} f_{J/\psi}^2\hskip6pt.
\end{equation} (The above expression requires neither relativistic
nor strong interaction corrections.  However,  one may use
$f_{J/\psi} =
\sqrt{12 / m_{J/\psi}} \, \psi(0)$,  where $\psi$ is the spatial wave
function,  to put the preceding equation in another often quoted
form.)  Using
$\Gamma(J/\psi
\rightarrow e^+ e^-) = 5.36$ keV \cite{PDG} we get

\begin{equation}    f_{J/\psi} = 3.11 f_{\pi} = 290 \hskip3pt {\rm
MeV}.
\end{equation}  Similarly for the $\psi^\prime = \psi(2S)$ and
$\Gamma(\psi^\prime
\rightarrow e^+ e^-) = 2.14$ keV \cite{PDG} we get

\begin{equation}    f_{\psi^\prime} = 2.15 f_{\pi} = 200 \hskip3pt
{\rm MeV}.
\end{equation} We also use $f_K =1.22 f_\pi$ \cite{PDG},  and the
value
$f_{K^*} = 1.67f_\pi = 156$ MeV quoted by Nardulli \cite{Nardulli}
based on the decay $\tau
\rightarrow K^* + \nu_\tau$.

We will give the actual decay amplitudes for  one case,  choosing
the $B \rightarrow J/\psi + K$ as our example.   The amplitude,
using the asymptotic distribution amplitude for the kaon, is

\begin{equation} M = M_a + M_{bc} + M_d = (-0.09+3.38i) +
(1.22+0.10i) - 0.26 = 0.87+3.48i.
\end{equation} One sees that the imaginary part is dominant,
especially in the amplitude squared.

Our numerical results for the branching ratios are in table
\ref{table4}. We note that the results are somewhat low compared to
the data,  although in the ballpark.  Since uncertainty estimates on
the determination of $f_B$ are large,  we give the results for the
case that $f_B = f_\pi$,  which is within the error bars of the lattice
gauge theory calculation.  There results a smaller
$\Lambda^\prime$ leading to no significant change in the
$B \rightarrow D\ {\rm or}\ D^*$ decays,  so that we do not need to
give a revised version of Table \ref{table1}.  However,  the decay
rates of the $B$ into charmonium do increase somewhat,  as may be
seen in Table \ref{table5}.  Precision measurement of the $B$ and
$D$ decays constants would help eliminate some of this ambiguity.
Note also the sensitivity to the kaon's distribution amplitude.

\section{Decays into light mesons}

The decay into light mesons is mainly governed by the factorable
``color unsuppressed'' diagrams shown in Fig.
\ref{mesdec},  and involves the CKM factor $V_{ub}$.  For
$B \rightarrow \pi \pi$ ,  the amplitude is \cite{Bdec},

\begin{equation} M= {3A \over \epsilon_B} \int_0^1 dy \,
{{(1-y)(1+y-2\epsilon_B)}
\over {y - 2\epsilon_B - i\eta}} \tilde \phi_{\pi}(y)
\equiv {3A \over \epsilon_B} I_1 \nonumber\\
\hskip4pt,
\end{equation}
 where A in this case is

\begin{equation} A={16\pi \over {9}} \alpha_s G_F V_{ub} V_{ud}
f_B f_{\pi}^2\hskip4pt.
\end{equation} We use $V_{ub} = 0.1 V_{cb} = 0.0045$ and other
parameters as in the charmonium section.

For the asymptotic distribution amplitude one gets

\begin{equation} I_1 = (1-2\epsilon_B) \left[
\ln\left(1-2\epsilon_B \over 2\epsilon_B \right) + i\pi \right] -
{1\over 2}
\; \stackrel{\epsilon_B\rightarrow 0}{\longrightarrow}
\; \ln{1\over \epsilon_B} + i\pi,
\end{equation} while the Chernyak-Zhitnitsky distribution
ampitude yields

\begin{eqnarray} I_1 &=& \; 5 \left\{
(1-2\epsilon_B)(1-4\epsilon_B)^2 \left[
\ln\left(1-2\epsilon_B \over 2\epsilon_B \right) + i\pi
\right] - {1\over 6} - 2 (1-2\epsilon_B)(1-4\epsilon_B) \right\}
\nonumber \\ & \stackrel{\epsilon_B\rightarrow
0}{\longrightarrow} &
\; 5 \left\{ \ln{1\over \epsilon_B} + i\pi \right\}.
\end{eqnarray}
 For typical finite $\epsilon_B$ this is about
$2 {1\over 2}$ times bigger in magnitude than the asymptotic
distribution amplitude result.

The decays into $\rho \pi$ or $\rho \rho$ are easy to describe in
the limit where we neglect the rho mass.  The formulas are the
same as those just quoted,  {\it mutatis mutandi}.  For example,  for
the decay
$B^0
\rightarrow
\rho^+ \pi^-$,  the $\rho$ comes from the decay of the virtual
$W^+$ and the only change is the replacement of one factor
$f_{\pi}$ by $f_{\rho}$.  As another example,  $B^0
\rightarrow \rho^- \pi^+$ requires replacement of the one decay
constant but also replacement of the pion distribution amplitude by
the rho's.

The distribution amplitude for the longitudinally polarized rho (the
only possibility in the massless rho limit) is given in \cite{CZZ} as

\begin{equation}
\tilde \phi_{\rho}(x) = 1.50 (2x-1)^2 + 0.70
\hskip6pt.
\end{equation} This could also be described as 30\% pion
Chernyak-Zhitnitsky distribution amplitude plus 70\% pion
asymptotic distribution amplitude,  allowing use of the pion
formulas already presented.  Incidentally,  no transversely
polarized rho's emerge in the massless limit because
 the relevant matrix element, eqn.
\ref{vectordecayconstant},  is proportional to the rho mass times
the polarization vector.  The massless limit is easily taken for
transverse polarization and easily gives zero. The longitudinal
polarization vector is singular in the massless limit,  so that the
limit is tricky and not zero.

Table \ref{table3} gives the decay rates of the
$B^0$ into $\pi$'s and $\rho$'s, in the massless $\pi$ and
$\rho$ limit and keeping just the factorable diagrams.  The reader
should have no trouble converting to other initial
$B$'s.  Again,  our model does not receive a significant contribution
from the color-suppressed diagrams.  The present upper limit for
the $B^0 \rightarrow \pi^+\pi^-$ branching ratio is $90 \times
10^{-6}$ and the corresponding limit for the $B^+$ decay is about
2$\frac{1}{2}$ times weaker
\cite{PDG}.

The actual decay rate of the $B$ into transverse $\rho$ pairs can be
estimated by keeping just the part of the transverse amplitude
linear in the rho mass.  For $B$ decay,  as a reflection of the
underlying behavior of the weak interaction,  only right handed
transverse rho's can emerge,  with the opposite being true for
$\bar B$.  The amplitude into transverse rho's is,

\begin{equation}  M = {6A \over \epsilon_B} {m_{\rho}\over m_B}
(1+\epsilon_B) \int_0^1 dy \, {{1-y}
\over {y - 2\epsilon_B - i\eta}} \tilde \phi_{\rho}(y)
\equiv {6A \over \epsilon_B} {m_{\rho}\over m_B} I_2
\hskip4pt.
\end{equation} This amplitude is smaller than that for the
longitudinal case by
$O(2m_{\rho} / m_B)$,  so we may anticipate a decay rate increase
of about 10\% by adding in the transverse contributions.  The
integral is

\begin{equation} I_2 = (1+\epsilon_B) \left\{(1-2\epsilon_B)
\left[
\ln\left(1-2\epsilon_B \over 2\epsilon_B \right) + i\pi
\right] - 1 \right\}
\end{equation} for the asymptotic wave function.  If we let
$\tilde \phi_{\rho}(y) \rightarrow 5(2y-1)^2$ we get

\FL
\begin{equation} I_2 = \; 5(1+\epsilon_B) \left\{
(1-2\epsilon_B)(1-4\epsilon_B)^2
\left[ \ln\left(1-2\epsilon_B \over 2\epsilon_B \right) + i\pi
\right] - {28\over 3}+{50}\epsilon_B -88\epsilon_B^2
\right\} ,
\end{equation} which we may use to evaluate a non-asymptotic rho
distribution amplitude.  The distribution amplitude for the
transverse rho need not be the same as for the longitudinal case,
and we shall only quote the numerical result for the asymptotic
case.  This appears in the last line of Table
\ref{table3},  which should be added to the penultimate line to get
the total branching ratio into~$\rho\rho$.

\section{Summary and Conclusions}

We have calculated decays of the $B$ meson into two-body final
states. The
$B$ is heavy and two-body final states guarantee maximal
momentum transfer.  We have used perturbative QCD in the belief
that the momentum transfer is sufficient to allow this.
Additionally,  we obtain the form factors for the $B \rightarrow
\bar D$ and $B \rightarrow \bar D^*$ currents which determine
the semi-leptonic decays;  our calculations here can be valid only
at the high momentum transfer end of their range.

Many of our results involve decays where factorable,  color
unsuppressed, diagrams (where both quark and antiquark from the
$W$ go into the same hadron) contribute.  The decays  $B
\rightarrow {\rm charmonium}+K^{(*)}$ are examples where
there are no factorable, color unsuppressed,  contributions.  The
factorable diagrams dominate in any process where they
contribute.  For the decays $B \rightarrow \bar D \pi$ we have
presented calculations of both the color unsuppressed and color
suppressed contributions,  and have shown that the color
suppressed contributions are suppressed dynamically also and are
smaller than the color unsuppressed contributions by more than
just a factor of~$N_c$.

For mesons involving at least one heavy quark we make a ``peaking
approximation,''  which is a statement that the wave function or
distribution amplitude is sharply peaked at a quark momentum
determined by the masses of the quark and of the bound state.  The
amplitudes and form factors we obtain from factorable diagrams
for the $B
\rightarrow \bar D$ or $B \rightarrow \bar D^*$ currents are
real.

Amplitudes  for decays involving a light meson which does not
come directly from the $W$ may have an imaginary part.  Examples
include $B
\rightarrow \pi \pi$ and $B \rightarrow J/\psi K$.  The imaginary
part comes when the integral over the light meson distribution
amplitude leads to a condition where one of the internal quark
propagators goes on shell.  The quarks despite being on-shell can
propagate only a short distance so that perturbative calculations
may still be used.  The imaginary parts are important because they
are often larger in magnitude than the real parts for the processes
where they contribute,  and it is they that make the $B$ decays into
$\pi \pi$ or $J/\psi K$ as large as they are.

We have calculated branching ratios for many two body $B$ decays
and presented them in sections and tables devoted to $B
\rightarrow \bar D^{(*)} +\rm another\ meson$,  to
$B \rightarrow {\rm charmonium}+K^{(*)}$,  and to $ B
\rightarrow {\rm light\ mesons}$.  For the first two categories,  we
have only calculated decays that are leading in CKM factors;  the
decays into light mesons are already subleading in CKM because
they involve $V_{ub}$ and we have calculated no decays that
involve a further CKM suppression.  Where there is data,  we are
often in agreement with it and always in the ballpark.

We feel that we have the right physics for explaining what is or
will be seen at the high momentum transfer end of $B$ decay.
Decays in addition to those we have already calculated could be
addressed.  In particular,  the decay $B
\rightarrow K^* + \gamma$ is amenable to our techniques and we
feel that the spectator quark has not properly been taken into
account in existing calculations.

{\underline{Acknowledgments}}:  We thank S. J. Brodsky, V. L.
Chernyak,  P. Rubin,  and J. D. Walecka for useful discussions.  This
work was supported in part by NSF Grant PHY-9112173 and DOE
Grant DOE-FG02-93ER-40762.

\figure{The weak decay of a pseudoscalar meson that determines
the
        normalization of the two quark Fock component in pQCD.
   \label{decay}}

\figure{The leading order color-unsuppressed diagrams, factorable
at the hadronic
       level,  contributing
       to the two body decay of the B meson.  Curly lines are gluons,
wavy
       lines are W bosons.
   \label{mesdec}}

\figure{Color-suppressed, ``internal W'' diagrams which can
contribute to $B^-
        \rightarrow D^0 \pi^-$ or $B^0$ decay into two neutral mesons.
   \label{cs1}}

\figure{Color-suppressed, ``W-exchange'' diagrams which can
contribute to
       $B^0 \rightarrow D^+ \pi^-$.
   \label{cs2}}

\figure{Reduced form factors for $B\rightarrow D\ {\rm or}\
D^*$.  The
      normal solid
       curve is for  $f_+$;  the dashed curve for $f_-$;  and the
markers indicate
       the form factors for the $B \rightarrow D^*$ with ``$\circ$'' for
$f$,
       ``$\triangle$'' for $g$, ``+'' for $a_+$,  and ``$-$'' for $a_-$.  Note
       that $q^2$ is the mass of whatever recoils against the
       $D^{(*)}$,  and that low $q^2$ corresponds to high recoil for the
       $D^{(*)}$
       and {\it vice versa}.  The upper limit of $q^2$ is at the zero
recoil point,
       and is 11.6 GeV$^2$ for the $D$ and 10.7 GeV$^2$ for the $D^*$.
The effect
       of the numerator corrections within just the
       $B \rightarrow D^*$ group are seen by comparing to the curve
for $g$,
       as the numerator  corrections have little effect upon $g$.  The
dispersion
       among the
       curves at low $q^2$ is a measure of the heavy quark symmetry
violation.
   \label{FF}}

\figure{Reduced form factors for $B\rightarrow D\ {\rm or}\ D^*$
plotted vs. $v
        \cdot v^\prime$.  Again,  the normal solid curve is for $f_+$;
the dashed
       curve for $f_-$;  and the markers are ``$\circ$'' for $f$,
``$\triangle$''
       for $g$, ``+'' for $a_+$,  and ``$-$'' for $a_-$.  The upper limit is
1.59 for
       the $D$ and 1.50 for the $D^*$.
   \label{FFom}}

\figure{Color-suppressed, non-factorable at the hadronic level,
       diagrams labelled for $B$ decays involving charmonium
   \label{BJdecay}}

\pagebreak

\begin{table}
\caption{Decay branching ratios as calculated from factorable
diagrams.  The models differ in the shape of the form factors,  and
also in the decay constants used.  In both cases,  we used $f_B = 1.42
f_\pi$, $f_D = 1.58 f_\pi$,  $f_\rho = 1.62 f_\pi$,
$f_{D^*} = 1.47 f_D$,  and we let $\alpha_s = 0.30$ and $f_{D_s^*} =
f_{D_s}$.  For the basic model (see text) we are led to $f_{D_s} =
f_\pi$ and for model II we used
$f_{D_s} = 1.7 f_\pi$.  The decays into $D\rho$, $D^* D_s$, $D^*
\rho$,  and the ratio of the two $D_s^*$ decays are predictions.}

$$\matrix{{\rm Branching\ ratios(\%)}&{\rm Basic\ model}&{\rm
Model\ II}&{\rm Data\ (for\ }B^0{\rm )}
\cr \quad B\rightarrow D \pi     \hfill & 0.32 & 0.32 & 0.32 \pm 0.07
\cr \quad B\rightarrow D D_s     \hfill & 0.77 & 0.97 & 0.90 \pm 0.60
\cr \quad B\rightarrow D \rho    \hfill & 0.90 & 0.82 & 0.80 \pm 0.50
\cr \quad B\rightarrow D D_s^*   \hfill & 0.62 & 0.65 & \cdots
\cr \quad B\rightarrow D^*\pi    \hfill & 0.31 & 0.31 & 0.32 \pm 0.07
\cr \quad B\rightarrow D^* D_s   \hfill & 0.58 & 0.68 & 0.80 \pm 0.40
\cr \quad B\rightarrow D^*\rho   \hfill & 1.04 & 0.94 & 1.60 \pm
1.10
\cr \quad B\rightarrow D^* D_s^* \hfill & 2.00 & 1.87 & \cdots}$$

\label{table1}
\end{table}

\begin{table}
\caption{Mass parameters for dipole fits to the form factors,  used
in ``Model II.''}
$$\matrix{{\rm Mass\ parameter} & {\rm Value\ (GeV)}
\cr M_+    & 5.09
\cr M_-    & 4.74
\cr M_f^*  & 5.11
\cr M_g^*  & 6.00
\cr M_+^*  & 5.25
\cr M_-^*  & 6.95 }$$

\label{table2}
\end{table}

\begin{table}
\caption{ Branching ratios for the indicated decays,  using $f_K =
1.22 f_\pi$,
$f_{K^*} = 1.67 f_\pi$, and neglecting both the mass of the $K$ and
of the $K^*$.  In this table,  $f_B = 1.42 f_\pi$,  $f_D = 1.58 f_\pi$,
$\Lambda^\prime = 545$ MeV,  and for charmonium decays
$\alpha_s = 0.35$.  The results using the pion distribution amplitude
for the kaon are mainly for informational purposes,  and we have
only used the asymptotic distribution amplitude in calculating
contributions from the transverse $K^*$.  Data for the
$K^*$ decay modes includes both longitudinal and transverse
polarizations.}
\centerline{
\begin{tabular}{lcccc}
 Branching ratio&\multicolumn{3}{c}{distribution amplitudes}&  \\
 (units $10^{-4}$) for&asymptotic&CZZ\cite{CZZ}&pion CZ
\cite{CZpi}&data (for$B^0$ \cite{PDG})   \\
\hline
$B^0\rightarrow\eta_c+K   $      & 0.8     & 0.8    & 2.2    &
$\cdots$            \\
$B^0\rightarrow\eta_c+K^* $      & 1.5     & 1.8    & 4.1    &
$\cdots$            \\
$B^0\rightarrow J/\psi+K$        & 1.2     & 1.4    & 2.5    &
$ 6.5\pm3.1 $       \\
$B^0\rightarrow J/\psi+K_L^*$    & 2.2     & 2.1    & 4.7    & $ 13.\pm
4.0 $      \\
$B^0\rightarrow J/\psi+K_T^*$    & 1.1     &$\cdots$&$\cdots$
&{\rm(part of above)} \\
$B^0\rightarrow \psi^\prime+K$   &0.7      &0.5     & 0.8   &$< 15.
$               \\
$B^0\rightarrow \psi^\prime+K_L^*$ & 1.2  & 1.1    & 1.5    &
$14.\pm 9.0$       \\
$B^0\rightarrow \psi^\prime+K_T^*$    & 1.2     &$\cdots$&$\cdots$
&{\rm(part of above)} \\
\end{tabular} }

\label{table4}
\end{table}

\begin{table}
\caption{Alternative branching ratios for the indicated decays,
using
$f_K = 1.22 f_\pi$,
$f_{K^*} = 1.67 f_\pi$, and neglecting both the mass of the $K$ and
of the $K^*$.  In this table,  $f_B = f_D = f_\pi$,  $\Lambda^\prime =
455$ MeV,  and for these charmonium decays
$\alpha_s = 0.39$.  Results using the pion distribution amplitude for
the kaon are mainly for informational purposes,  and we used only
the asymptotic distribution amplitude in calculating rates for the
transverse $K^*$.  Data for the $K^*$ decay modes sums
longitudinal and transverse polarizations.}

\begin{tabular}{lcccc}
 Branching ratio&\multicolumn{3}{c}{distribution amplitudes}&  \\
(units $10^{-4}$) for&asymptotic&CZZ\cite{CZZ}&pion CZ
\cite{CZpi}&data (for$B^0$
\cite{PDG})  \\  \hline

$B^0\rightarrow\eta_c+K   $      & 1.1     & 1.4    & 3.0    &
$\cdots$
\\
$B^0\rightarrow\eta_c+K^* $      & 2.1     & 2.5    & 5.6    &
$\cdots$
\\
$B^0\rightarrow J/\psi+K$        & 1.3     & 2.1    & 3.9    &
$ 6.5\pm3.1 $      \\
$B^0\rightarrow J/\psi+K_L^*$    & 2.4     & 2.4    & 7.3    & $ 13.\pm
4.0 $     \\
$B^0\rightarrow J/\psi+K_T^*$    & 1.3   & $\cdots $& $\cdots $
&{\rm(part of above)} \\
$B^0\rightarrow \psi^\prime+K$   & 0.7   & 0.8    &  1.5 &$<
15.$                \\
$B^0\rightarrow \psi^\prime+K_L^*$ & 1.2  & 1.2  & 2.8  & $14.\pm
9.0$      \\
$B^0\rightarrow \psi^\prime+K_T^*$ & 1.1   & $\cdots$ & $\cdots$
&{\rm(part of above)} \\
\end{tabular}

\label{table5}
\end{table}

\begin{table}
\caption{ Branching ratios for the indicated decays.}

\centerline{
\begin{tabular}{lcc}
 Branching ratio&\multicolumn{2}{c}{distribution amplitudes}\\
  for (units $10^{-6})$&asymptotic&CZ(\cite{CZpi}) or CZZ(\cite{CZZ})
\\
\hline
$B^0\rightarrow\pi^+\pi^- $      & 0.7 & 3.5 \\
$B^0\rightarrow\rho^+\pi^-$      & 1.8 & 9.2 \\
$B^0\rightarrow\pi^+\rho^-$      & 1.8 & 2.8 \\
$B^0\rightarrow\rho_L^+\rho_L^-$ & 4.9 & 7.3 \\
$B^0\rightarrow\rho_T^+\rho_T^-$ & 0.5 &$\cdots$\\
\end{tabular}
 }

\label{table3}
\end{table}

\newpage

\vglue 1.4in  
\hskip 1.5in {\special{picture DKconst scaled 1000}}
\centerline{Fig. \ref{decay}.}
\vskip 0.1in

\vglue 2.0in  
\hskip 0in {\special{picture mesdec scaled 1000}}
\centerline{Fig. \ref{mesdec}.}
\vskip 0.1in

\vglue 3.0in  
\hskip 0.4in \special{picture intW scaled 1000}
\centerline{Fig. \ref{cs1}.}
\vskip 0.1in

\vglue 3.0in  
\hskip 0.4in \special{picture Wexch scaled 1000}
\centerline{Fig. \ref{cs2}.}
\vskip 0.2in

\vglue 3.5in  
\hskip 0.4in \special{picture BdecayFF scaled 1000}
\centerline{Fig. \ref{FF}.}
\vskip 0.2in

\vglue 3.0in  
\hskip 0.4in \special{picture BdecayFFo scaled 1000}
\centerline{Fig. \ref{FFom}.}
\vskip 0.3in

\vglue 4.0in  
\hskip 0in \special{picture BJdecay scaled 1000}
\centerline{Fig. \ref{BJdecay}.}
\vskip 0in


\begin{references}
\bibitem{comment2} As in
\cite{Bdec}, we think of ``factorable" as meaning factorable at the
hadron level,  i.e., both quark and antiquark from the decay of the
$W$ go into the same hadron.  However, other authors use
``factorable'' to mean just non-involvement of the spectator in the
weak interaction. We have tried to be clear in the present work by
using ``color-(un)suppressed'' as an extra modifier when needed.
\end{references}
\end{document}